\documentclass[11pt,amsmath,amssymb]{article}
\usepackage{amsmath}

\newcommand{\beq}{\begin{equation}}
\newcommand{\eeq}{\end{equation}}
\newcommand{\bit}{\begin{itemize}}
\newcommand{\eit}{\end{itemize}}
\newcommand{\ben}{\begin{enumerate}}
\newcommand{\een}{\end{enumerate}}
\newcommand{\la}{\langle}
\newcommand{\ra}{\rangle}

\newcommand{\ch}{{\cal{H}}}

\newcommand{\co}{{\cal{O}}}
\newcommand{\bs}{\boldsymbol}
\newcommand{\f}{\frac}

\newcommand{\mb}{\mbox}

\newtheorem{criterion}{Homogeneity Criterion}

\newcommand{\be}{\begin{equation}}
\newcommand{\ee}{\end{equation}}
\newcommand{\bea}{\begin{eqnarray}}
\newcommand{\eea}{\end{eqnarray}}

\begin{document}
%\doublespacing
\begin{titlepage}

\begin{flushright}
\today
\end{flushright}

\vspace{1in}

\begin{center}

{\bf Modeling Quantum Mechanical Observers via Neural-Glial Networks}

\vspace{1in}

\normalsize

%\large
{Eiji Konishi\footnote{E-mail address: konishi.eiji@s04.mbox.media.kyoto-u.ac.jp}}

\normalsize
\vspace{.5in}

 {\em Faculty of Science, Kyoto University, Kyoto 606-8502, Japan}

\end{center}

\vspace{1in}

\baselineskip=24pt
\begin{abstract}We investigate the theory of observers in the quantum mechanical world by using a novel model of the human brain which incorporates the glial network into the Hopfield model of the neural network. Our model is based on a microscopic construction of a quantum Hamiltonian of the synaptic junctions.
Using the Eguchi-Kawai large $N$ reduction, we show that, when the number of neurons and astrocytes is exponentially large, the degrees of freedom of the dynamics of the neural and glial networks can be completely removed and, consequently, that the retention time of the superposition of the wave functions in the brain is as long as that of the microscopic quantum system of pre-synaptics sites.
Based on this model, the classical information entropy of the neural-glial network is introduced. Using this quantity, we propose a criterion for the brain to be a quantum mechanical observer.
\end{abstract}

\vspace{.7in}
 
\end{titlepage}
\section{Introduction}
Today, in the community of quantum physicists, the standard interpretation of quantum mechanics is based on the theory of the Copenhagen school.\cite{Dirac,Pauli,QM1,QM2} This states that, first, each quantum system (in the following we just write {\it{system}}) is described by a complex-valued wave function which obeys unitary time development as a solution of the Schr${\ddot{{\rm{o}}}}$dinger equation (causal and continuous development) and, second, in the measurement of an observable of a system, the wave function $|\psi\ra$ of the measured system collapses in a non-unitary time development to an eigenfunction $|a\ra$ of the observable with the probability $|\la a|\psi\ra|^2$ (stochastic interpretation). However, the Copenhagen interpretation contains the famous measurement problem concerning the paradoxical consequences of a complex system composed of a measured quantum system and its measurement system. The time development of the wave function due to the measurement in this system is unitary and non-unitary, which is contradictory. This is due to an ambiguity in the borderline between the concepts of measured systems and their measurement systems that involves the definition of observers. A resolution of this problem can be seen, for example, in von Neumann's infinite regress of measurements, that introduces the observer as an abstract ego and assumes the projection hypothesis such that the process of reading a datum in the regression by the abstract ego completes the measurement process.\cite{vNeumann1,vNeumann2,Wigner} Nevertheless such the abstract egos have not been part of the study of physics.

\renewcommand{\theenumi}{\roman{enumi}}
\renewcommand{\labelenumi}{\theenumi)}

Fifteen years ago,  R. Penrose and S. R. Hameroff proposed a scenario describing observer systems, which in this paper are human brains, to be a clue to the resolution of this unsatisfactory feature of the Copenhagen interpretation.\cite{HP1,HP2} Among their ideas, the one relevant to our study can be summarized as follows.
\ben
\item The conscious activities of the human brains contain a non-computable and non-algorithmic process.
\item As far as is known, a candidate for such a process is the collapse of a superposition of wave functions, and they adopt it.
\item Quantum gravity effects concerning the fluctuation of time increments cause the objective reduction of wave functions.
\een 
When we generalize their statement i) to include the singular measurement property of the observers, their description of quantum mechanics does without the concept of an abstract ego. The measurement activity of the abstract ego, that is, the projection hypothesis, is replaced by the quantum fluctuation of the time increment.
Then, the system of the quantum mechanical world plus quantum mechanical observers can be seen in total as an object of study in quantum physics. A brief account of the physics part of this thesis by Penrose\cite{Penrose} is given in Appendix A.

 Penrose and Hameroff associated the location of the state reduction, which is recognized as a conscious activity, with the microtubules in the brain. This is because the quantum states of the microtubules are able to be macroscopically coherent due to the ordered forms of the microtubules and thus have a long enough retention time of the superposition of the wave functions (decoherence time).
On the other hand, based on today's consensus between brain scientists, which has much experimental support, it is necessary to include the functional system consisting of spike activities with their threshold structure in any model of the brain which, directly or indirectly, explains consciousness.\cite{Kandel} Besides this point, the coherent assemblies of neuronal cells are also observed in neuroscience, not just the spike activity of single cells.\cite{Freeman1,Freeman2}
 Thus, we apparently need to revise their original model to use the fields of spike activities and the coherent activities of assemblies of neuronal cells.

In this paper, we pursue the Penrose-Hameroff scenario of the quantum theory of observers by considering a model based mainly on the neural and glial networks rather than microtubules. In this model we invoke the mechanism of the Eguchi-Kawai large $N$ reduction that originated in lattice gauge theory\cite{EK}. Such a model has never been discussed in the original literature\cite{HP1,HP2}. Here, the neural network is based on a microscopic construction of a quantum Hamiltonian of the synaptic junctions according to Ricciardi-Umezawa theory, which will be reviewed in Section 2.\cite{RU1,RU2,RU3} Throughout this paper, we refer to the network of astrocytes, which are one of the glial cells in the human brain, as the {\it{glial network}}.

The goal of this paper is to understand why this singular property of observers, that is, quantum state reduction, occurs without any ambiguity in the definition of the observers whose activities consist of spikes, by showing that the retention time of a superposition of quantum states in the neural-glial system is as long as that of the pre-synaptic sites. The borderline between measured systems and measurement systems is defined by the classification of quantum systems according to their decoherence time.

Our main physical subject is, as will be explained in the next section, a globally coherent quantum field theoretical ground state in the brain, which describes as classical fields a macroscopically steady electro-magnetic field and the macroscopic electric dipole field of water molecules (i.e., a field describing a macroscopic number of the electric dipoles that have a coherent direction and strength) caused by ferroelectric and hydrophilic materials such as cell membranes and the dendrites of neurons. We will show that this ground state describes a network of Josephson currents over the whole of the brain where the coherent regions are bridged by microtubules and other cytoskeletal structures as a superradiative circuit through non-coherent regions.\cite{JHHPY,JPY} From this ground state, for $n$ neurons, we express the excitatory or inhibitory neural states, and their junctions proportional to the expectation values of the polarization currents of the ions between post- and pre-synaptic site junctions, which couple to neural states in the Hopfield type Hamiltonian\cite{H1,H2}.

We assume a homogeneity criterion for the dynamics of the phase of the neural state (in the neural-glial system), as will be explained in Sections 3 and 5. Due to this criterion and the assumption of the existence of the glial modulation of the synaptic transmission, we can model the neural-glial networks so that there exist gauge symmetries with a rank of the order of the logarithm of $n$. If this rank times the time span of the function of the brain, counted by spikes, is {\it{large}} (in a qualitative sense, as it is used in the original Eguchi-Kawai large $N$ reduction argument in elementary particle physics\cite{EK}), all of the classical degrees of freedom (d.o.f.) are reduced. Consequently, the retention time of the superposition of wave functions of the brain, in the sense of the Penrose thesis, is globally retained as long as the behavior of each pre-synaptic site and the quantum coherence property is still consistent with the global functions of the brain which consist of spike activities and the coherent activities of large masses of neurons and glia cells\cite{Freeman1,Freeman2}.  As will be explained in Section 2, the coherence property of the pre-synaptic site is related to that of the microtubules, which can be recognized as cylindrical wave guides of the coherent photons\cite{JHHPY,GDMV1}.

As a consequence, we can explain the physical origin of the observers that make measurement processes in the quantum mechanical world description by invoking the Penrose thesis about state reduction\cite{Penrose}.

The structure of the rest of this paper is as follows. In Section 2, we give a review of the quantum field theoretical model of the brain based on the works by Jibu, Pribram and Yasue and Vitiello.\cite{JPY,V} In Section 3, based on the ground state in the brain given in Section 2, we give a quantum field theoretical derivation of the neural network model. In Section 4, we incorporate the glial network into the neural network based on its function of maintaining the homeostasis of the concentrations of ions, neurotransmitters and water in the synaptic gaps. In Section 5, we give a condition under which the Eguchi-Kawai large $N$ reduction works. In the final section, we discuss the quantum theoretical definition of observers.
\section{Ricciardi-Umezawa Theory: A Review}
Many decades ago, L. M. Ricciardi and H. Umezawa hypothesized the existence of  interacting quantum field d.o.f. in the human brain and interpreted the codes of memories as the vacuum expectation values of macroscopic coherent quantum fields, for instance, Bose-Einstein condensate and laser under the assumption that the Hamiltonian possesses spontaneously broken symmetries.\cite{RU1,RU2,RU3} They made this proposal to explain efficiently the {\it{stability}} and {\it{non-locality}} properties of memory due to these properties of the ground state in quantum field theory. From the exhaustive study by Jibu, Pribram and Yasue\cite{JPY}, which is based on the papers by Fr${\ddot{{\rm{o}}}}$hlich and the early papers about the role of dipole wave quanta in living matter\cite{GDMV1,Froehlich1,Froehlich2,GDMV2,GDMV3}, these interacting macroscopic quantum fields are identified, in the framework of quantum electrodynamics, with the electromagnetic field and the electric dipole fields of the ordered water molecules around the ferroelectric and hydrophilic materials such as cell membranes and dendrites of neurons. A notable point of their theory is that it is based on the quantum field theoretical description of macroscopic scale (micro-meter order) objects. Such a description is possible due to the infiniteness of the d.o.f. of the quantum fields and quantum mechanics with finite d.o.f. cannot describe them except at the microscopic scale (less than nano-meter order).

In this section, we review the Ricciardi-Umezawa theory in order to construct the quantum field theoretical ground state in the brain. We use thermal dissipative quantum d.o.f. to describe the electric dipole fields of water molecules and of the polarizations of the intracellular and extracellular ions (Na$^+$, K$^+$, Cl$^{-}$ and Ca$^{2+}$) which have their ion channels in neurons or astrocytes. To simplify the situation, our model of the brain assumes that the kinds of dipole field are unique.

The following description is based on thermo field dynamics (TFD)\cite{U,Umezawa} since the considering system is a dissipative one. The TFD formalism necessary to describe the dissipative nature of the brain dynamics has been firstly introduced by Vitiello's paper\cite{V,Vbook}, where the dissipative quantum model of brain was first formulated as an extention to the dissipative dynamics of the Ricciardi-Umezawa model in order to cure the very small capacity of memory of that model. In the framework of TFD, one introduces bosonic annihilation and creation operators for the original modes and for mirror modes which are {{independent}} from the original modes
\begin{equation}
 {{a}}_\kappa\;,\ \ {{a}}^\dagger_\kappa\;,\ \ \tilde{{{a}}}_\kappa\;,\ \ \tilde{{{a}}}^\dagger_\kappa\;,\label{eq:dual}
\end{equation}
where the tilde conjugation $\sim$ satisfies
\begin{subequations}
\begin{align}
&(A_1A_2)^\sim=\tilde{A}_1\tilde{A}_2\;,\ \ (c_1A_1+c_2A_2)^\sim=c_1^\ast \tilde{A}_1+c_2^\ast \tilde{A}_2\;,\\& (\tilde{A})^\sim=A\;,\ \ (A^\dagger)^\sim=\tilde{A}^\dagger\;,
\end{align}
\end{subequations} 
for arbitrary operators $A_1$, $A_2$ and $A$ and $c$-numbers $c_1$ and $c_2$. The mirror modes are introduced in order to describe mixed states of a single quantum d.o.f. as pure states of a double quantum d.o.f. formed from the original and its mirror. Here $\kappa$ is a label for the d.o.f. of quanta, for example, spatial momentum.
These operators satisfy the bosonic canonical commutation relations (CCR) \begin{equation}[{{a}}_\kappa,{{a}}_\lambda^\dagger]=\delta_{\kappa\lambda}=[\tilde{{{a}}}_\kappa,\tilde{{{a}}}_\lambda^\dagger]\;,\ \  [a_\kappa,\tilde{a}_\lambda]=0=[a_\kappa^\dagger,\tilde{a}_\lambda^\dagger]\;.\end{equation}

In TFD,\cite{U,Umezawa} the thermal expectation value $\bar{A}$ of any observable $\hat{A}$ is rewritten in the form of an expectation value of it in the ground state $|0(\beta)\rangle$, that is, the zero energy pure eigenstate of the free part of the Hamiltonian of the system for temperature $T=1/(k_B\beta)$ as\begin{equation}\bar{A}=\langle0(\beta)|\hat{A}|0(\beta)\rangle\;.\end{equation}
The time development of the ground state $|0(\beta)\rangle$ is given by the dissipative Schr${\rm{\ddot{o}}}$dinger equation:
\begin{equation}
i\hbar\frac{\partial }{\partial t}|0(\beta)\rangle=\hat{{\cal{H}}}_{tfd}|0(\beta)\rangle\;,
\end{equation}
where the total Hamiltonian of the quantum system of the electromagnetic field and the electric dipole fields of the ordered water molecules and of the polarizations of the intracellular and extracellular ions in the perimembranous region of the neurons, which is tildian (i.e., $i$ times it is invariant under the tilde conjugation), can be split into two terms, \begin{equation}\hat{{\cal{H}}}_{tfd}=\hat{{\cal{H}}}_{dq}+\hat{{\cal{H}}}_{em}\;.\label{eq:totalH}\end{equation}
As shown later, the interaction term in $\hat{{\cal{H}}}_{em}$ spontaneously breaks the rotational symmetry of the dipole field. We denote the creation and annihilation operators of the dipole field and dipole wave quanta (Goldstone bosons) by $\sigma_\kappa^\dagger$ and $\sigma_\kappa$, and by $\chi^\dagger_\kappa$ and $\chi_\kappa$, respectively.
 
In order to rewrite the dissipative system in a clean form,\cite{Dis} we define the operators \begin{equation}\Sigma_\kappa=\frac{\sigma_\kappa+\tilde{\sigma}_\kappa}{\sqrt{2}}\;,\ \ \tilde{\Sigma}_\kappa=\frac{\sigma_\kappa-\tilde{\sigma}_\kappa}{\sqrt{2}}\;,\ \ X_\kappa=\frac{\chi_\kappa+\tilde{\chi}_\kappa}{\sqrt{2}}\;,\ \ \tilde{X}_\kappa=\frac{\chi_\kappa-\tilde{\chi}_\kappa}{\sqrt{2}}\;,\end{equation} and their canonical conjugates $\Sigma_\kappa^\dagger$ and $\tilde{\Sigma}_\kappa^\dagger$, and $X_\kappa^\dagger$ and $\tilde{X}_\kappa^\dagger$. The new operators satisfy the bosonic CCR\begin{subequations}\begin{align}
&[{{\Sigma}}_\kappa,{{\Sigma}}_\lambda^\dagger]=\delta_{\kappa\lambda}=[\tilde{{{\Sigma}}}_\kappa,\tilde{{{\Sigma}}}_\lambda^\dagger]\;,\ \  [\Sigma_\kappa,\tilde{\Sigma}_\lambda]=0=[\Sigma_\kappa^\dagger,\tilde{\Sigma}_\lambda^\dagger]\;,\\
&[{{X}}_\kappa,{{X}}_\lambda^\dagger]=\delta_{\kappa\lambda}=[\tilde{{{X}}}_\kappa,\tilde{{{X}}}_\lambda^\dagger]\;,\ \  [X_\kappa,\tilde{X}_\lambda]=0=[X_\kappa^\dagger,\tilde{X}_\lambda^\dagger]\;.\end{align}\end{subequations}

The free part of the Hamiltonian of the dipole field in the perimembranous region of the neurons is
\begin{equation}
\hat{{\cal{H}}}_{dq}=\sum_\kappa \hbar \Omega_\kappa(\Sigma^\dagger_\kappa \Sigma_\kappa-\tilde{\Sigma}^\dagger_\kappa\tilde{\Sigma}_\kappa)\;,
\end{equation}
where $\Omega_\kappa$ is the frequency of the dipole field quantum (dq).

The Hamiltonian of the electromagnetic fields in the perimembranous region of the neurons consists of a free part and an interaction part,\cite{JPY} \begin{equation}\hat{{\cal{H}}}_{em}=\hat{{\cal{H}}}_{free}+\hat{{\cal{H}}}_{int}\;.\end{equation}
 The free part is\begin{equation}\hat{{\cal{H}}}_{free}=\sum_\kappa\sum_{p=1,2}\hbar \omega_\kappa ({{a}}_{\kappa,p}^{\dagger}{{a}}_{\kappa,p}-\tilde{{{a}}}_{\kappa,p}^{\dagger}\tilde{{{a}}}_{\kappa,p})\;,\label{eq:EM}\end{equation}where $\omega_\kappa$ is the frequency of photons and ${{a}}_{\kappa,p}^{\dagger}$ and ${{a}}_{\kappa,p}$ are the creation and annihilation operators of photons with polarization $p$.
The Hamiltonian for the interaction between radiative photons and dipole field quanta is
\begin{equation}
\hat{{\cal{H}}}_{int}=\sum_\kappa\sum_{p=1,2}ig({{s}}_{+,\kappa} {{A}}^{\dagger}_{\kappa,p} -{{s}}_{-,\kappa}{{A}}_{\kappa,p})\;, \label{eq:damp}
\end{equation}where $g$ is coupling constant and we have defined ${{A}}_{\kappa,p}={{a}}_{\kappa,p}\tilde{{{a}}}_{\kappa,p}$ and
\begin{equation}
{{s}}_+={{\Sigma}}^\dagger \tilde{{{\Sigma}}}^\dagger\;,\ \ {{s}}_-={{\Sigma}}\tilde{{{\Sigma}}}\;,\ \ {{s}}_3=\frac{1}{2}({{\Sigma}}^\dagger {{\Sigma}}+\tilde{{{\Sigma}}}^\dagger\tilde{{{\Sigma}}}+1)\;.
\end{equation}
The operators ${{s}}_+$, ${{s}}_-$ and ${{s}}_3$ form an $su(1,1)$ algebra due to the bosonic CCR for the dipole fields.

As a result of the interaction in $\hat{{\cal{H}}}_{int}$, the rotational symmetry of the dipole field is broken.
Simultaneously, the electromagnetic fields also acquire non-zero vacuum expectation values.
By recognizing our system as a quantum dissipative one of the damped Goldstone bosons, as performed in Ref.23, the Hamiltonian $\hat{{\cal{H}}}_{int}$ can be rewritten as the one for the dissipation of dipole wave quanta:
\begin{equation}
 \hat{{\cal{H}}}_{dis}=\sum_\kappa i\hbar \Gamma_\kappa (X^\dagger_\kappa\tilde{X}^\dagger_\kappa-X_\kappa \tilde{X}_\kappa)\;,\end{equation} where ${\Gamma}_\kappa$ is the damping constant of the dipole wave quantum, which is formed from the coupling constant $g$ and the vacuum expectation values of electromagnetic fields.\cite{Dis}

The ground state $|0(\beta)\rangle$ satisfies Glauber's condition for a coherent wave function that carries an $SU(1,1)$ group factor:\cite{G}
\begin{equation}
|0(\beta)\rangle=\exp({-iG(\vartheta)})|0(\beta)\rangle_0\;,\ \ G(\vartheta)=-i\sum_\kappa\vartheta_\kappa(X^\dagger_\kappa \tilde{X}^\dagger_\kappa-X_\kappa\tilde{X}_\kappa)\;,\label{eq:ground}
\end{equation}
with
\begin{equation}
X_\kappa|0(\beta)\rangle_0=\tilde{X}_\kappa|0(\beta)\rangle_0=0\;.
\end{equation}
The representations of two vacua and the CCR of the operators $(X^\dagger,X,\tilde{X}^\dagger,\tilde{X})$ are unitarily inequivalent in the infinite volume limit, if and only if the sets of eigenvalues of the particle number operators of Goldstone bosons and those of the mirror modes \begin{equation}{\cal{N}}_{X_\kappa}=X_\kappa^\dagger X_\kappa\;,\ \ {\cal{N}}_{\tilde{X}_\kappa}=\tilde{X}_{\kappa}^\dagger\tilde{X}_\kappa\;,\end{equation} are not equal.
The shifts between unitarily inequivalent representations of CCR may be associated with the imprinting of memory, since both processes are accompanied by the violation of time reversal symmetry. The foliations of the wave function under such processes can possess infinitely many distinguishable codes ${\cal{N}}$. Under time development by $\exp(-i\hat{{\cal{H}}}_{dis}t/\hbar)$ the quantity ${\cal{N}}_{X_\kappa}-{\cal{N}}_{\tilde{X}_\kappa}$ is a constant of motion since \begin{equation}[\hat{{\cal{H}}}_{dq},\hat{{\cal{H}}}_{tfd}]=0\;,\end{equation} holds. We note that these codes ${\cal{N}}$ have an $SU(1,1)$ group factor since \begin{equation}{\cal{N}}_{X_\kappa}={\rm{sinh}}^2\vartheta_\kappa\;.\end{equation} This mechanism was first discovered to be a natural consequence of dissipative quantum field theory by Vitiello.\cite{V}
 Due to this mechanism in the dissipative system, the capacity of memory codes, which are defined in the sense of the Ricciardi-Umezawa proposal, is as large as the number of varieties of trajectories of time development of the wave function, and can be infinite.\cite{V}

 The ground state of the system, whose spatial localization is the set of normalized quantum field states $|{{C}}_i\rangle$ of the quanta of the electric dipole field, photons and Goldstone bosons in the perimembranous region of the synaptic site of the $i$-th neuron $C_i$, $i=1,2,\ldots,n$, has the structure
\begin{equation}|0(\beta)\rangle= \phi_i|{{C}}_i\rangle\ \ {\mb{in}}\ \ C_i\;,\ \ i=1,2,\ldots,n\;,\label{eq:state}\end{equation}
for real-valued coefficients ${{\phi}}_i$. The phase factor of $\phi_i|C_i\rangle$ is absorbed into $|C_i\rangle$.
(Of course, there is an ambiguity of a factor $\pm1$ in the phase. This factor plays a very important role in our model, as will be explained in the next section.) This ground state $|0(\beta)\rangle$ stores memories as its order parameters in the spontaneous symmetry breakdown.\cite{RU1,RU2,RU3,V}
The results in this section can be summarized by the structures of the ground states in Eq.(\ref{eq:state}).
 Each quantum state $|C_i\rangle$ may be macroscopically coherent due to the general collective behavior of the dipole field quanta of water molecules as a laser and the spontaneous breakdown of the rotational symmetry of the ground state of quantum dipole fields around ferroelectric and hydrophilic materials such as cell membranes and dendrites of neurons.\cite{JPY,water} Here, we use the term {\it{quantum field state}} highlight the fact that we are not talking about the classical field. The {\it{classical field}} is defined by the criterion \begin{equation}\f{{\Delta {{n}}}}{{{{n}}}}\ll 1\;,\end{equation} where $n$ is the particle number and $\Delta{{n}}$ is its quantum fluctuation.\cite{TFD}

 Actually, from the exhaustive study by Jibu, Pribram and Yasue of the dynamics of water molecules in the perimembranous region of the neurons, it has been shown theoretically that Bose-Einstein condensates of evanescent photons with high enough critical temperature exist. The evanescent photons are generated by absorbing the Goldstone mode into the longitudinal mode of the radiation field as in the Higgs mechanism.\cite{JPY,GDMV1} Moreover, at each synaptic site the superconducting currents across the Josephson junction (i.e., Josephson currents), that produces a quantum tunneling between two such macroscopically quantum coherent regions of dendrites, whose coherence lengths are about $50$ micrometers, separated by a thin enough quantum incoherent region, are theoretically predicted.\cite{JPY} Between the pre- and postsynaptic cells is a gap about 20 nanometers wide\cite{Kandel}, and the unit of the dendritic net falls within these coherence lengths. Consequently, in the brain a global tunneling circuit exists\cite{JPY} and in this sense the global nature of Eq.(\ref{eq:state}) is satisfied.\cite{GDMV1,GDMSV} We quote the original arguments from their paper.
  
  \smallskip
  \smallskip
  
   ``{\it{..... As those boson condensates of evanescent photons are directly related to the macroscopic quantum dynamics of the radiation field, certain superconducting phenomena could take place there. Indeed, the longitudinal mode of the radiation field plays the role of the order parameter characterizing the macroscopic dynamics of superconducting media, because it is locked to the phase of any matter field with electric charge through the gauge transformation.}}
   
    {\it{Recall that the dendritic membrane is composed of two oppositely oriented phospholipid molecules. Thus, not only the outer layer provide for hydrophilic extracellular processing, but the inner layer also makes possible an ordered water medium within the dendrites (and their spines).}}
    
    {\it{Consequently, we can expect that, within the patch (or compartment) of dendrite (including its spine) that falls within the coherence length of the ordered water, a couple of outer and inner perimembranous regions separated by a thin layer of cell membrane form a Josephson junction, that is, a sandwich-structured junction of two superconducting region, weakly coupled with each through the membrane by means of quantum tunneling mechanism. .....}}''
    
\smallskip
\smallskip

We make a few comments on the relevance of the microtubules to the Ricciardi-Umezawa theory. In the perspective proposed by Hameroff,\cite{Hameroff} the microtubules are thought to act like dielectric waveguides for photons, that is, quantum dynamical modes of an electromagnetic wave. Based on this perspective, in the paper by Jibu et al\cite{JHHPY}, the microtubules and other cytoskeletal structures are theorized to play the roles of non-linear coherent optical devices by a quantum mechanical ordering phenomenon termed by {\it{superradiance}} with characteristic times much shorter than those of thermal interaction.\cite{GDMV1} We quote the original consequent arguments from the paper by Jibu et al.\cite{JHHPY}

\smallskip
\smallskip

``{\it{..... Superradiance and self-induced transparency occuring in ordered water within the hollow core of cylindrical microtubules behaving as waveguides will result in coherent photons. This coherence, estimated to be capable of superposition states among microtubules spatially distributed over hundreds of micrometers, which in turn are in superposition with other microtublues hundreds of micrometeres away in other directions and so on, could account for a coupling of microtublues dynamics over wide areas. This in turn could account for a unity of thought and consciousness. .....}}''

\smallskip
\smallskip

In the final section, in order to estimate the decoherence time of the individual pre-synaptic site, we will refer to this argument.

In the quantum field description, the real parts of the $c$-number coefficients $\phi$ of the localized wave functions in Eq.(\ref{eq:state}) correspond to the neuron states, also denoted by $\phi$, in the semi-classical model Eq.(\ref{eq:start}). Because of the reduction of the dynamical d.o.f. of spikes by Eq.(\ref{eq:EK}) shown later as our original argument, it is obvious that the informational representations of spike activities in the brain depend on the quantum field states $|{{C}}_i\rangle$ as well as the Shannon representation of the bits of spikes in the brain.\cite{Shannon}

\section{Derivation of the Neural Network}
In this section, we derive the neural network of the spike activities from the quantum field theory of the Ricciardi-Umezawa framework.

Today, the definite formulation of spike activities is based on the Hodgkin-Huxley model.\cite{HH} This models the cell membrane and ion-channels of a neuron by the condenser and dynamical registers in an electric circuit. The voltage-dependent sodium (Na$^{+}$) and potassium (K$^{+}$) ion channels are embedded in neuronal cell membranes and keep the equilibrium electric potential by adjusting the ion concentrations inside and outside of the neurons. Each voltage-dependent ion channel has probability factor for its opening. This circuit obeys simple non-linear differential equations for the conservation of electric currents via the electric potential and inflowing currents. This depolarization of membrane potentials induced by a sodium ion current greater than a threshold value is termed a spike. After the generation of spikes, the membrane potential repolarizes and returns to the resting state by the inactivation of the sodium channel and the activation of the potassium channel.

The mechanism of the emission of the neurotransmitters is as follows.\cite{Kandel}
When a spike (i.e., a depolarization) arrives at the pre-synaptic site, the voltage-gated calcium channels open. Then, the outer calcium ions (Ca$^{2+}$) flow into the pre-synaptic site, and due to the action of these calcium ions, a vesicle will couple to the pre-synaptic membrane. Then, the neurotransmitters in the vesicle are emitted into the synaptic gap. The number of emitted neurotransmitters is proportional to the concentrations of calcium ions in the pre-synaptic site and the time span of the opening of the voltage-gated calcium channels.

The dynamics of spikes in the Hodgkin-Huxley theory essentially consists of non-linear oscillations. This activity is compatible with the Hopfield model\cite{H1,H2} of the neural network in a statistical mechanical fashion and can be encoded in its discrete variables. The neural network models are recognized as models of learning and associative memory. Here, {\it{associative memory}} indicates that the system will settle down to stable patterns of the excitatory neurons (i.e., the memories in these models) which are determined by the types of the inputs. There are various neural network models of learning with synaptic plasticity: for example, the multilayer perceptron model and the mutually coupled model.\cite{Hopfield} In mutually coupled models, such as the Hopfield model, learning processes quadratically strengthen the excitatory couplings of neurons by the synaptic plasticity for the patterns of memories embedded in these couplings, according to the Hebbian learning rule. Unlearning processes strengthen the inhibitory couplings by random patterns.\cite{Crick,dream} In this paper, to make a reasonable simplification of our arguments, we will not study the learning functions of the neural network.

 Since the neural network system can be regarded as a non-linear electric circuit of spikes, the energy of this circuit is given by the summation of the products of the expectation value of the electric charge density of the ions mediating spikes and the post synaptic potentials over all of the synaptic sites. However, the ground state constructed in the last section is effective only in the perimembranous regions of the synaptic sites.
So, we pay attention to the fact that the intracellular and extracellular ions, and the water molecules in the perimembranous regions of the synaptic sites\cite{JPY}, create currents that are proportional to the post synaptic potentials in the brain. Using this clue, we make a new attempt to construct the neural network Hamiltonian from quantum field theory.

The Hamiltonian of the neural network, which is added to Eq.(\ref{eq:totalH}), is given by products of the expectation value of the electric charge of 
the ions mediating spikes, the electric synaptic resistance and the operator of the polarization current in the perimembranous regions of the synaptic sites:
\begin{equation}\hat{\ch}_{net}=\f{1}{2}\sum_\kappa Q\hbar({\boldsymbol{n}}\cdot  \kappa) (\Sigma_\kappa^\dagger \Sigma_\kappa-\tilde{\Sigma}^\dagger_\kappa \tilde{\Sigma}_\kappa)\;,\label{eq:net}\end{equation}where the time dependent total electric charge of the ions mediating spikes in the neural network is denoted by $Q$, and ${\boldsymbol{n}}$ is the unit vector field of the transverse directions of the axons at the synaptic junctions from one neuron to another. The orientation of ${\boldsymbol{n}}$ is given by the sign of the electric charges of the messenger ions in the excitatory or inhibitory synapses. In Eq.(\ref{eq:net}), we assume two simplifications. First, the density of the polarization dipoles reflects that of the ions mediating spikes. In the present model, we simplify this reflection to be an identity relation.
Second, the expectation values of Eq.(\ref{eq:net}) reflect the number of calcium ions (${\rm{Ca}}^{2+}$) flowing in through the voltage-dependent calcium channels at the pre-synaptic sites, and reflect indirectly the synaptic inputs on the other neurons via the neurotransmitters. In the present model, we simplify these reflections to be proportionality relations. For convenience, we select the unit of electric charge and the unit of resistance to be those of the unit dipole and synaptic resistor.

In the neural network model, we define memories in the quantum field description by incorporating the definition in the Hopfield model, that is, Hebb's law on the plasticity of the synapses, in the sense that both of them assign the role of memories to the strengths $J$ of neuron junctions in the Hamiltonians.\cite{H1,H2} In the quantum field description, we replace the classical values of $J_{ij}$ by the time dependent expectation values of the quantum neural network Hamiltonians $\hat{{\cal{H}}}_{net}^{j\to i}$ 
\begin{equation}J_{ij}=-2\langle {{C}}^{j\to i}_j|\hat{{\cal{H}}}^{j\to i}_{net}|{{C}}^{j\to i}_i\rangle\;,\label{eq:memory}\end{equation} on the common domains $C^{j\to i}$ of two neurons $C_{i}$ and $C_j$ at the synaptic junction of $i$-th neuron where the expectation values of $\hat{{\cal{H}}}^{j\to i}_{net}$ do not vanish.  Since the variables $|C^{j\to i}\rangle$ and $\phi_i$ are independent of each other, $J_{ij}$ also is independent of $\phi_i$. We note that the definition of memories in Eq.(\ref{eq:memory}) is descended from both the Hopfield theory and the Ricciardi-Umezawa theory.

For the neural states $\phi_i$ in Eq.(\ref{eq:state}) and the junctions $J_{ij}$ between them, with $i,j=1,2,\ldots,n$, the vacuum expectation value of the Hamiltonian of the neural network in Eq.(\ref{eq:net}) has a Hopfield form:
\begin{equation}
\ch_{hop}=-\f{1}{2}\la \phi,J\phi\ra\;,\label{eq:Hop}
\end{equation}
where $\la A,B\ra$ denotes the inner product of $A_i$ and $B_i$ on their index $i=1,2,\ldots,n$, and in analogy with the time development in the Hopfield model, the presence or absence of a spike of the $i$-th neuron is expressed as $\phi_i>0$ or $\phi_i<0$, respectively. Due to Eq.(\ref{eq:state}), the $n$-dimensional vector $\phi_i$ and the wave function Eq.(\ref{eq:state}) are normalized, i.e., \begin{equation}\sum_i\phi_i^2=1\;.\label{eq:norm}\end{equation}This condition is equivalent to the normalization of the wave function $|0(\beta)\ra$. 

When a set of neurons ${\cal{S}}$ is given, the time developments of the phases $\vartheta_i[k+1]$ of the neural states $\phi_i[k+1]$, whose absolute value is defined by Eq.(\ref{eq:state}), are ruled to be given by the recursion equations:
\begin{equation}\phi_i[k]=e^{i\vartheta_i[k]}\hat{\phi}_i[k]\;,\ \ \vartheta_i[k]\in\{0,\pi\}\;,\ \ \hat{\phi}_i[k]\in{\bs{R}}_{\ge0}\;,
%\phi_i[k]=e^{i\vartheta_i[k]}\bigl||0(\beta)\ra_{C_i}[k]\bigr|\;,\ \ i=1,2,\ldots,n\;,
\label{eq:brainwave}\end{equation}with
\begin{eqnarray}e^{i\vartheta_i[k+1]}={\mb{sign}}\Biggl(-\sum_jE_{ij}e^{i\vartheta_j[k]}\Biggr)={\mb{sign}}\Biggl(\sum_j J_{ij}\phi_j[k]\Biggr)\;,\label{eq:ev}\end{eqnarray}due to actually $\hat{\phi}>0$, where $[k]$ with $k=1,2,\ldots,N-1$ represents the temporal steps and $\phi[1]=\phi$ and $E_{ij}$ is the synaptic part of the potential energy of the synaptic junction between two neurons $C_i$ and $C_j$\begin{equation}
E_{ij}=-\f{1}{2}\hat{\phi}_i\hat{\phi}_jJ_{ij}\;.
\end{equation} In Eq.(\ref{eq:ev}), the corresponding threshold potential energy value to produce a spike at the post-synaptic site of $i$-th neuron is given by $-\sum_jE_{ij}$ and may be time dependent. Here, we recall that the number of emitted neurotransmitters is proportional to the concentration of calcium ions in the pre-synaptic site. Since the calcium channel is voltage dependent (here, we must not confuse the active membrane potential with the potential of the synaptic current),\cite{Kandel} we can simplify the model so that the concentrations of calcium ions are common between all pre-synaptic sites of a neuron without losing the physical essence of the model. Due to Eq.(\ref{eq:ev}), we can treat $\phi$ as the dynamical variables.
In Eq.(\ref{eq:brainwave}), $e^{i\vartheta_i}$ expresses whether there is a spike or not by $+1$ or $-1$ respectively and $\hat{\phi}_i$ describes the semi-classical behavior of 
%the dipole fields reflecting that of 
the ions mediating spikes and gives the expectation value of their electric charge density.
We define the vector of the temporal set of neural states
\begin{equation}
(\varphi[k])_i={{\phi}}_i[k]\;,\ k=1,2,\ldots,N\;,
\end{equation} for a total number of temporal steps $N$. We refer to these temporal steps of the redefinitions of the neural states as {\it{renormalizations}}.

In Eq.(\ref{eq:ev}), the time developments consist of two parts. The first part is the quantum mechanical unitary or non-unitary transformations of the ground state $|0(\beta)\ra$ and the second part is the non-linear threshold time development on $e^{i\vartheta}$ in the conventional model of the neural network.
The dynamical d.o.f. of our neural network model is that of temporal transformations indexed by $[k]$, $k=1,2,\ldots,N$.

 Regarding the phase factor of the neural states $e^{i\vartheta}$, we assume the following homogeneity criterion on the neural state dynamics.
 \begin{criterion}The $n$ distinct configurations of the sites of neuron $i$ can be represented by the signs of their renormalized states. Namely, \begin{equation} {\cal{S}}\simeq\{ {\rm{sign}}(\varphi_i[k])|\ i=1,2,\ldots,n\;,\ k=1,2,\ldots,N\}\;,\label{eq:hom}\end{equation}holds.\end{criterion}
 In the neural network (not in the neural-glial network) it may not hold exactly but only in a weaker form. This criterion first requires that in any pair of neurons its elements have different time developments of $e^{i\vartheta}$. Besides this condition, this criterion requires the periodicity condition on neural dynamics. In Section 5, this criterion will be generalized for generalized neural states (see Eq.(\ref{eq:general})) and its second condition will be expressed as the integrability of the neural-glial system. Thus, by considering the glia's physiological functions, which will be explained in the next section, the second condition of this criterion is natural in the neural-glial system. Throughout this paper, to simplify our arguments, we assume that the dynamics of the system considered is constrained to satisfy this criterion exactly by initial conditions. Practically, we assume the first condition of this criterion.

 Due to this assumption, the site information of neurons is coded in an $N$-dimensional vector of signs of neuron states. The total number of steps is \begin{equation}N=\lfloor\log_2n\rfloor\;,\label{eq:N}\end{equation}where $n=\dim\phi+1$ and Gauss' symbol is defined by \begin{equation}x-1<\lfloor x\rfloor\le x\;,\ \ \lfloor x\rfloor \in{\boldsymbol{Z}}\;.\end{equation}
We will revisit this criterion in Section 5 after we take into account the glial degrees of freedom.

\section{Neural-Glial Network}
During the past two dacades, a revolution has occurred in the recognition of the functions of astrocytes.\cite{Glia1,Glia21,Glia22,Glia23,Glia24} Our new statistical model of the neural and glial network takes into account this new view.

The astrocytes have recognized to have mainly three functions from the physiological view point:\cite{Glia24} the modulation of the synaptic transmission, the neural synchronization and the regulation of cerebral blood flow. Between them, the one of interest here is focused on the first function including the maintenance of the homeostasis of the concentrations of ions, neurotransmitters and water. We explain it through the following three processes.\cite{Glia22,Glia24}
 
 \renewcommand{\theenumi}{\alph{enumi}}
\renewcommand{\labelenumi}{(\theenumi)}

\begin{enumerate}
\item As has been recently discovered, each astrocyte communicates with the others through gap junctions via the calcium ion (Ca$^{2+}$) wave produced by the calcium-induced calcium release from the intracellular calcium stores of astrocytes.\cite{wave1,wave2,wave3,wave4,wave5} Consequently, the activation of a glial receptor by release of neurotransmitters (glutamate) from the pre-synaptic site results in the modulation of distant synapses by release of neurotransmitters (glutamate and ATP) from other astrocytes via the calcium ion waves. This means that the modulation of synaptic junctions (see (b)) among different synapses is done globally.
\item The neurotransmitters (glutamate) flowing at the synaptic sites are modulated by the astrocytes\cite{Mod1,Mod2,Mod3,Mod4,Pot1,Pot2}. The activation of glial receptors of astrocytes by the release of neurotransmitters (glutamate) from the pre-synaptic site, where this release is evoked by every spike event, inputs to the intracellular concentrations ${\mb{In}}(t)$ into the calcium store at a time $t$, and the output to the extracellular concentrations of the neurotransmitters (glutamate) of astrocytes ${\mb{Out}}(t)$, that is, the activation of the pre- and post-synaptic receptors with the regulation of the synaptic transmitter release can be modeled to satisfy
\begin{equation}
{\mb{Out}}(t)=C^{(1)}{\mb{In}}(t)\;,\ \ {\mb{Out}}(t)=\sum_{k=1}^MC^{(2)}_k(t){\mb{Out}}(t-(k-1)t_0)\;,
\end{equation} for constant $C^{(1)}$, time interval of spikes $t_0$\footnote{For the simplicity of the model, we model the time intervals between spikes to be constant.} in the order of 1ms to 100ms\cite{Kandel} and time span $Mt_0$ of the modulation, without losing the physical essence of the model. In$(t)$ and Out$(t)$ are vectors at every time, with indices corresponding to the synapses, and $C^{(1)}$ and $C^{(2)}_k(t)$ are matrices. These modulations maintain the homeostasis of the concentrations of ions, neurotransmitters and water in the synaptic gaps.\cite{Maintain} So, there are constraints on $C^{(2)}_k(t)$.
\item
When the astrocytes bridge different synapses via their calcium ion waves and modulate them, the glial action describes the feedforward and feedback properties of the regulation of pre-synaptic junctions. These properties are due to the cyclic activation, via the modulation by astrocytes, of more than one synaptic junction, such as heterosynaptic depression and the potentiating of inhibitory synapses etc.\cite{Pot1,Pot2}
\end{enumerate}

To define the model of the astrocytes mathematically, we consider two points. First, the energy of the system (see Eq.(\ref{eq:start})) always tends to decrease towards the minimum. Second, the glia's function of the maintenance of the homeostasis (a) and (b) and the feedback or feedforward type of their modulations (c), denoted by ${\cal{G}}$, means that the glial outputs ${\mb{Out}}(t)$, which are temporally accumulated in the synaptic gaps by the temporal changes of the neural states, are linearly averaged in a time range\footnote{This is because the glial network has no threshold structure.\cite{Glia1}} \begin{equation}I_0=t_0\times [1,M]_{\bs{N}}\;,\end{equation} by the linear transformation $\exp({\cal{G}})$ on the corresponding $M$ inputs via the temporal neural state vector $\varphi$. We note that this linear transformation should be recognized not as the modulation of the neural states at different times but the modulation of the glial outputs ${\mb{Out}}(t)$ to the synapse accumulated in the synaptic gaps at different times. Due to (c), these outputs reflect the future inputs by the feedback or feedforward property of the glial modulations. (Here, based on the glia's function of the maintenance of homeostasis (a), that is, the globality of the astrocyte action on the synapses and (b), the glial actions ${\cal{G}}$ are defined to be Lie algebra valued, with a linear basis that is related by Noether's theorem to the modes of their modulations as constants of motion in the modulation of the synaptic junctions. Due to arguments that will be explained soon, we assume that this variable ${\cal{G}}$ takes its value in the orthogonal Lie algebra $o(M)$.) The time range $I_0$ determines the time span of the maintenance of homeostasis. In this paper, to simplify the case, we assume that the time range is unique. Our modeling is more advanced than averaging by adding a kinetic term to the Hamiltonian since the latter approach does not incorporate such a time span. Let us assume this time range $I_0$ is equal to the one of the periodic cycles of the neural states in the neural-glial system (see Eq.(\ref{eq:general})), that is, \begin{equation}M=N\;.\label{eq:main}\end{equation}
This assumption is under the following logic. First, if the periodic cycles of the neural states in the neural-glial system $N$ exist, their unit time span is longer than the time span of the maintenance of homeostasis due to the definition of the latter, $M\le N$. Second, for the latter time span $M$, since such the gilal action makes the neural system tend to be linear, approximately $N\le M$. Then, at least under this approximation for the second logic, Eq.(\ref{eq:main}) holds.

By keeping in mind the temporal decrease of the value of the Hamiltonian, under the two-fold structure with Eq.(\ref{eq:Hop}), we define the Hamiltonian written using the bilinear form of $\varphi$ to be
\begin{equation}\ch=-\f{1}{2N}\la\la\varphi,\exp(\Delta) \varphi\ra\ra\;,\label{eq:start}\end{equation}
 where $\la\la A,B\ra\ra$ denotes the inner products of $A_i[k]$ and $B_i[k]$ by contracting on both $i=1,2,\ldots,n$ and $k=1,2,\ldots,N-1$. We have also introduced the covariant difference 
\begin{equation}\Delta\varphi=\delta\varphi+{\cal{G}}\varphi\;,\label{eq:gliaren}\end{equation}
with 
\begin{equation} \delta\varphi[k]=\varphi[{k+1}]-\varphi[{k}]\;.\end{equation}
 By analogy with the dynamics of the Ising model at zero temperature, we find from Eq.(\ref{eq:start}) a recursion equation for the generalized neural state vector $\hat{\varphi}$ of the neural-glial system, which no longer satisfies the normalization condition,
 \begin{equation}
\hat{\varphi}[1]=\varphi[1]\;,\ \ \hat{\varphi}[k+1]=\exp(\Delta)\hat{\varphi}[k]\;,\ \ k=1,2,\ldots,N-1\;.\label{eq:general}
 \end{equation}
When we normalize $\hat{\varphi}$, it is interpreted as the same expression of the neural state by $\varphi$ and will be used to classify the non-linearities of the neural-glial system in Section 5.
  The glial variable ${\cal{G}}$, whose full condition is \begin{equation}{\cal{G}}_{kl}\in o(N)\;,\label{eq:gliadef}\end{equation} has indices $k$ and $l$ representing time values. We assume that the span $I_0$ represents a periodic pattern to retain the homogeneity criterion. Namely, the summation $\sum_l{\cal{G}}_{kl}\varphi[l]$ over $l=1,\ldots,N$ for a resulting index $k$, which is actually $k^\prime+pN$ for a natural number $k^\prime \in (I_0/t_0)$ and a natural number $p$, means the summation of past elements over the time range $t_0\times [k^\prime+(p-1)N+1,k^\prime+pN]_{\bs{Z}}$. We note that Eqs. (\ref{eq:Hop}) and (\ref{eq:start}) represent different physical systems. The latter system is larger than the former system by the number of degrees of freedom of the astrocytes.

The glia's activity is originally defined in the infinite time span and irrelevantly to the periodic time span $N$ of the neural states at that stage of Section 3. However, due to Eq.(\ref{eq:main}), it is related to $N$ and the constants of motion associated to the basis of $o(N)$ mean that ${\cal{G}}$ is the unit of the periodic pattern in the $o(\infty)$ matrix.

 We remark on the relation of Eq.(\ref{eq:start}) to the neural network Eq.(\ref{eq:Hop}). For the unitary or non-unitary time promotion operator $\hat{U}$ of the states in Eq.(\ref{eq:state}), using the relation between the orthogonal matrices with the sizes $n$ and $N$ as will be seen in Eq.(\ref{eq:vee}), the replacement 
  \begin{equation}\delta\to \Delta\;,\label{eq:LTP0}\end{equation} in Eq.(\ref{eq:gliaren}) is recognized as a temporal redefinition of $\hat{U}$ and $e^{i\vartheta}$ that absorbs the new degrees of freedom of the astrocytes. We note that, due to the rule Eq.(\ref{eq:ev}), this replacement reflects the glia's function on the modulation of synaptic transmission and $\exp(\delta)$ is sufficient to represent the coupling of neural states with the $J$ matrix.

Owing to the homogeneity criterion Eq.(\ref{eq:hom}), our model Eq.(\ref{eq:start}) has {{local}} $O(N)$ symmetry transformations ${\cal{O}}_r$, whose spatial variables are defined to be the signs of the $N$-vector $\hat{\varphi}$, on the index $k$ (the renormalization steps):\begin{eqnarray}\varphi\to {{\cal{O}}}_r\varphi\;,\ {\cal{G}}\to {\cal{O}}_r{\cal{G}}{\cal{O}}_r^{t}-(\delta {\cal{O}}_r) {\cal{O}}_r^t\;,\  {{\cal{O}}}_r\in O(N)\;,\label{eq:sym}\end{eqnarray}
where $\delta \co_r$ is the variation of the dependence of $\co_r$ on the site of neuron by the variation of the neuron site $\delta \varphi$ defined by Eq.(\ref{eq:hom}). The generators of this gauge symmetry are related to the constants of motion in the glial modulation of the synaptic junctions. Due to the local property of the symmetry, the glial variable ${\cal{G}}$ is recognized as a gauge field.

This gauge symmetry of Eq.(\ref{eq:sym}) in Eq.(\ref{eq:start}) means that, due to the modulation of synaptic junctions, the distinction between the neurons corresponding to the variable sites of the local symmetry transformations $\co_r$ by the time process labeled by the index $[k]$ loses its validity.

The reasons why we model the glial gauge group to be $O(N)$ and construct the symmetry transformations in Eq.(\ref{eq:sym}) are as follows.
 Due to Eq.(\ref{eq:norm}), the neural states $\phi[k]$ are $O(n)$ vectors.
Then, the $O(n)$ global transformation $\co$ on the neural state is also a transformation of the Hamiltonian in Eq.(\ref{eq:start}) for an $O(N)$ matrix $({{\cal{O}}}_r)_i$ with the local index $i=1,2,\ldots,n$, which corresponds to the site of $i$-th neuron, such that the $O(N)$ vector part (renormalization part) of the variables $\varphi$ satisfies the identity\begin{equation}{\cal{O}}\varphi={{\cal{O}}}_r\varphi\;.\label{eq:vee}\end{equation}Eq.(\ref{eq:vee}) which is explicitly written as 
 \begin{eqnarray}
\sum_j{\cal{O}}_{ij}\phi_j=\sum_l({\cal{O}}_r)_i^{kl}\phi_i[l-k+1]\;,\ \ i=1,2,\ldots,n\;,\label{eq:Or}\end{eqnarray}for $k=1,2,\ldots,N$ has at least one solution ${{\cal{O}}}_r$ except for the $N=1$ case, since against $n$ equations there are $n\f{N(N-1)}{2}$ degrees of freedom. The solutions of Eq.(\ref{eq:Or}) are distinct if the vectors $\phi[k]$, for $k=1,2,\ldots,N$, are linearly independent.

It should be noted that the idea of gauging the synaptic connection in classical and quantum neural network models was initially proposed and investigated in three papers\cite{Matsui1,Matsui2,Matsui3}. However, we note that our context for the gauge symmetry is independent of theirs and our approach is new in the sense of that we incorporate the glial network into our model of the neural network based on the recently discovered roles and activities of the astrocytes.\cite{Glia1,Glia21,Glia22,Glia23,Glia24}
\section{Homogeneity Criterion on the Neural-Glial Network}
To extend the homogeneity criterion of the neural network to the neural and glial networks, we start with the definition of the classical information entropy of the neural and glial networks. Here, the extension is done by the replacement of $\varphi$ in Eq.(\ref{eq:hom}) by $\hat{\varphi}$ in Eq.(\ref{eq:general}).

In the neural network model, the spike of the neuron has been considered as a bit of information, and by itself we can consider a closed information structure. This is seen in the Boltzmann machine-type neural network model\cite{Boltzmann}, which is an extension of the Hopfield-type neural network model incorporating stochastic processes. However, when we take into consideration the glial network, it is more natural to define the information entropy using the interaction between the neuron states ${{\phi}}_i$ and the glia state ${{{\cal{G}}}}$. We introduce the classical entropy of the neural and glial networks by 
\begin{equation}
H(t)=-\sum_{s\in I(t)}\int\int d\varphi_s d{\cal{G}}(p_{t}(s t_0)\log_2 p_{t}(st_0))\;,\label{eq:cl}
\end{equation}
where $t$ labels the time evolution of the neural states $\varphi$ by a certain threshold structure as seen in the non-linear Hopfield model, $t_0$ is the unit of time (the interval between spikes) and $s$ indexes the time value.
In Eq.(\ref{eq:cl}), we define the set $I(t)$ of $s$, which is associated with the time $t$, as
\begin{equation}
I(t)={{\boldsymbol{Z}}_{\ge0}}\cap\biggl[0,\frac{t}{t_0}\biggr]\;.\label{eq:sequence}
\end{equation}
 The time distribution of the modulation of synaptic transmission by astrocytes is given by the probability
\begin{equation}
p_{t}(st_0)=\frac{1}{Z_{t}} \exp\biggl(\beta\frac{1}{2}\langle{{\varphi}}_s,(\exp(\Delta){{\varphi}})_s\rangle \biggr)\;,\label{eq:p}
\end{equation}
with 
\begin{eqnarray}
\sum_{s\in I(t)}\int\int d\varphi_sd{\cal{G}}(p_{t}(st_0))=1\;.\label{eq:sample}
\end{eqnarray}
The reason why we adopt $p_{t}(st_0)$ as the probability is that $\exp(\beta(1/2)\langle {{\phi}},J{{\phi}}\rangle)$ is the probability of the time evolution of the Boltzmann machine-type neural network\cite{Boltzmann}
 and it is generalized to Eq.(\ref{eq:p}) by the generalization of the synaptic junction $J$ to the glial action $\exp({{\Delta}})$.
In Eq.(\ref{eq:p}), the normalization factor $1/Z_{t}$ is the inverse of the partition function of the neural-glial system.
 The renormalization index $k$ is replaced by $s$ in Eq.(\ref{eq:sequence}), and the renormalizations are made by the time evolutions of Eq.(\ref{eq:ev}).

 We classify the non-linear behavior of neural-glial network using the classical information entropy in Eq.(\ref{eq:cl}). When
\begin{equation}
 H( t)\sim \log_2{ t}\;,\label{eq:period}
\end{equation} the classical system is linear or periodic (i.e., integrable); otherwise it shows chaotic behavior. Simultaneously, the extended homogeneity criterion on the neural-glial network holds.
 Here, we use a relation for almost all trajectories of time evolution with time variable $t$ and the probability $p_{t}(st_0)$ on the sample space of the neural and glial states in Eq.(\ref{eq:sample}):\cite{K}
 \begin{equation}
 H( t)\simeq K(t)\;,\label{eq:algo}
 \end{equation}
 where $K$ is the algorithmic complexity of a trajectory over a time $t$. $K$ is the length of the smallest program able to reproduce the trajectory on a universal classical Turing machine.\cite{K1,K2}

 For the property of being an observer, if the system satisfies the criterion in Eq.(\ref{eq:period}), we define an index $N$ which takes into account the time span of the brain function being considered:
  \begin{equation}N\equiv \biggl(\frac{\Delta t}{t_0}\biggr)\times \lfloor\log_2n\rfloor\;,\end{equation} and if 
 \begin{equation}N\gg1\;,\label{eq:measurement}\end{equation} the dynamical d.o.f. for the brain function, which does not appear till we observe it during a time span $\Delta t$, is reducible by the Eguchi-Kawai large $N$ reduction.\cite{EK} 

The statement of the Eguchi-Kawai large $N$ reduction is that, if we assume a large number of local symmetry generators (of course the Eguchi-Kawai large $N$ reduction is not valid for a global symmetry) and the existence of unbroken $U(1)$ phase symmetries between the gauge fields and their Hermite conjugates (in our case this latter assumption is not necessary), then the spatial d.o.f. can be completely removed from the partition function of the system due to the factorization properties\cite{FAC} of the loop correlation functions. We apply this statement to our model Eq.(\ref{eq:start}). We define the $O(N)$ vector part $\varphi_N$ of the $O(Nn)$ vector $\f{1}{\sqrt{N}}\varphi$, which is obtained by quenching the other degrees of freedom. Then, the functional integrals over the quenched neural state variable $\varphi_N$ and the glial variable ${\cal{G}}$ in the partition function are reduced, in the large $N$ limit, to matrix integrals over the $O(N)$ matrix $\Phi$ and the glial $O(N)$ matrix $\Gamma$, due to the relation tr$(\varphi^t\co \varphi)$=tr$(\co \varphi\varphi^t)$:
\begin{eqnarray}Z[\beta,J]
%&&
=\int\int {{D}}\varphi_N{{D}}{\cal{G}} e^{-\beta{\mathcal{H}}[\varphi_N,J,{\cal{G}}]}
%\nonumber\\&&
=\int\int d\Phi d\Gamma e^{-\hat{\beta}{\rm{Tr}}{\mathcal{H}}_{mat}[\Phi,J,{\Gamma}]}\;,\label{eq:EK}\end{eqnarray}
where $\ch_{mat}$ is the Hamiltonian of the reduced matrix model corresponding to Eq.(\ref{eq:start}) and $\hat{\beta}=\beta/N$.
 This means that, for the variable $\varphi$, the number of dynamical d.o.f. is reduced from $O(n^N)$ to $O(Nn)$: the sites of neurons change from being variables to being merely indices. The other d.o.f. of the $O(Nn)$ vector still survive. In the Eguchi-Kawai large $N$ reduction, the value ${\hat{\beta}}$ is kept constant.\cite{EK}
 Here, we have used the thermal variable $\hat{T}=1/(k_B{\hat{\beta}})$.
The Eguchi-Kawai large $N$ reduction in Eq.(\ref{eq:EK}) leads to the globalization of both the quantum correlations of the operators and the quantum mechanical properties of the neurons at their pre-synaptic sites.
\section{Summary and Discussion}
%\subsection{The Role of Human Brains as Observers}
As explained in the Introduction, in quantum mechanics, the measurement of any observable induces a non-unitary time development of a quantum system --- the collapse of the superposition. As seen in von Neumann's infinite regress of measurement processes, to describe the concept of measurement of observables, we require an {\it{observer}}. Now, we can define what an observer is. Here, we invoke Penrose's state reduction thesis, which claims that the non-unitary processes of measurement result from the quantum variance of the increment of time $\Delta t$ due to quantum fluctuations caused by the effects of quantum gravity.\cite{Penrose}
We denote the decoherence time of the neuron's pre-synaptic site and the brain's spatial domain $D$ (not of the superradiative circuit but of the neural network) by $\tau_{ps}$ and $\tau_{br}$, respectively. Then, our scheme for defining an observer is simply
\begin{equation}
\tau_{br}\sim \tau_{ps}\neq0\;,\label{eq:Penrose}
\end{equation}
even though the volume of the domain $D$ belongs to the classical limit of the wave function of each pre-synaptic site. Eq.(\ref{eq:Penrose}) is compatible with the functions of the neural network.
We note that Eq.(\ref{eq:Penrose}) does not always imply that there is a macroscopic superposition of the brain wave functions $|0(\beta)\ra$.

On the basis of the scheme in Eq.(\ref{eq:Penrose}), an observer would become just a quantum system in
which the superposition of the wave functions is maintained during a non-zero time span as well as in the microscopic system, and in which memories are the vacuum expectation values of order parameters of its wave function.

In the following, we briefly explain the scenario of the completion of a measurement process by an observer, which is expected from this scheme and the idealized roles of the neural network.

We assume an objective quantum system with a superposition of $l$ wave functions, which are the eigenstates of an observable $\hat{\co}$ with eigenvalues $\Lambda_i$ for $i=1,2,\ldots,l$. In contrast to an ordinary quantum system, the real human brain can recognize each eigenvalue of the observable $\Lambda_i$ in the informational database of neural state configurations, denoted by $\Phi_i$ for $i=1,2,\ldots,l$. 

First, by the brain's recognition, the information about $\hat{\co}$ would be translated into the information of the bits of spikes in the neural network. This process needs to be done between the superposed quantum states of the objective quantum system and the observer, since the classical informational mediation in the brain takes too long to make the collapses of the quantum superpositions coincide. The superposition of $|0(\beta)\ra$ corresponding to that of the objective quantum system will be generated by a unitary transformation on $|0(\beta)\ra$ via Eq.(\ref{eq:brainwave}). 

Second, the superposition of the wave functions of the objective quantum system would collapse due to the quantum variance of the time increment. Simultaneously, due to the common quantum variance of the time increment, the superposition of $|0(\beta)\ra$ would collapse within a wide enough time span $\tau_{br}$. (If the time span $\tau_{br}$ were vanishingly short, the coincidence of the collapses of the superpositions of the objective quantum system and the observer would be a rare occurrence.) Here, we recognize that {\it{free will}}, constrained by the probability law of the state reduction, works. Then, the neural state configuration, as a coefficient of the superposition of $|0(\beta)\ra$, would be chosen from $\Phi_i$, $i=1,2,\ldots,l$.

 Consequently, within the observer's conscious experience, the observer and the objective quantum system would enter the same world branch. The reason why their world branches are same is that the stochastic variable is not the wave function but the time increment $\delta t$. The results of the measurement processes would be recorded in the memories $J$. Due to its expression in Eq.(\ref{eq:memory}), $J$ is determined by the foliation of the collapsed branches of $|0(\beta)\ra$.

This scenario of the completion of measurement processes by a human brain summarizes our studies so far on the logical level. However,
to close this paper,
 we need to point out also that there are unfinished issues which are beyond the scope of the present paper and can be regarded as separate from our principal original ideas. As will be seen in what follows, there has already been some valid and original research concerning these issues.

First, to complete this scenario for the completion of measurement processes, there are more aspects to be considered. Such an aspect, we raise the conditions of the state reduction of the quantum state of each pre-synaptic site, which may be the same as the original microtubule scenario of Penrose and Hameroff\cite{HP1,HP2} since, as mentioned in Section 2, the role of microtubules as the wave guides of the photons in a superradiative circuit in the brain was theorized, so, the decoherence time of the pre-synaptic site is estimated to be equal to that of the microtubules. In the paper by Hagan et al\cite{HHT}, which is the response to the criticism by Tegmark\cite{Tegmark}, the decoherence time of the microtubules is calculated to be in the order of 10ms to 100ms due to the ordering of water around microtubule bundles. From this result, we confirm that the quantum theory may be still relevant to real conscious activities.
 Besides this aspect, numerical simulations are also required to understand the reflections of the initial data of the neural-glial system and the coherences of the pre-synaptic sites to the state reduction of the brain wave function according to our model. A numerical simulation of the decoherence time of the microtubules has been done by Hiramatsu et al.\cite{HMS}

Second, besides the role of the completion of measurement processes of physical quantities as explained above, there may be the high-order roles of the state reduction on intellectual brain activities, 
%(such as the refined concept of the free will in the conscious activities etc),%
 which have been discussed in Penrose's celebrated articles.\cite{Penrose1,Penrose2} 
 In this sense, our present theory of the quantum mechanical observers is primitive. However, since the purpose of the present theoretical investigation is to explain the physical definition of observers, in the present paper we do not discuss these details. As explained in the Introduction, the essential role is that the state reduction can be associated with the non-computable and non-algorithmic activities of the brain.

 I hope to take up some of the issues in future.
 
 \begin{appendix}
 \section{Brief Account for the Penrose Thesis}
 In this appendix, we present a brief account for the Penrose thesis on the role of the effects of quantum gravity on the state reduction.\cite{Yasue}
In order to find the concrete form of the statement of the Penrose thesis, for a Hamiltonian $\hat{{\cal{H}}}$ and the wave function $\psi({\boldsymbol{x}},t)$, we rewrite the inverse of the derivative by time $t$ in the Schr${\ddot{{\mbox{o}}}}$dinger equation 
\begin{equation}
i\hbar\frac{\partial \psi({\boldsymbol{x}},t)}{\partial t}=\hat{{\cal{H}}}\psi({\boldsymbol{x}},t)\;,
\end{equation} as an average over a normal stochastic variable $\delta t$:
\begin{equation}
\langle\psi({\boldsymbol{x}},t)\rangle=\exp\biggl(-\frac{it}{\hbar} \hat{{\cal{H}}} -\frac{\sigma t}{2\hbar^2}\hat{{\cal{H}}}^2\biggr)\langle\psi({\boldsymbol{x}},0)\rangle\;,\label{eq:est}
\end{equation}
where the average is defined by the following recursion equation
\begin{equation}
\langle \psi({\boldsymbol{x}},\mu)\rangle=\int dt^\prime\exp\biggl(-\frac{i\delta t^\prime}{\hbar}{\hat{\cal{H}}}\biggr)f(\delta t^\prime)\langle\psi({\boldsymbol{x}},0)\rangle\;.\label{eq:ave2}
\end{equation}
In Eq.(\ref{eq:ave2}) we make an average over a normal stochastic variable $\delta t$ with quantum variance $\sigma$, mean $\mu$ and distribution function $f(\delta t^\prime)$. In the text, we often refer to the second exponential factor in Eq.(\ref{eq:est}) as the {\it{quantum variance of the time increment}} $\delta t$.

In quantum mechanics, a Hamiltonian $\hat{{\cal{H}}}$ is a Hermitian operator. Thus for the eigenvalues $\{\lambda\}$ of $\hat{{\cal{H}}}$, there exists a unique spectral family $\{d\hat{{\cal{H}}}(\lambda)\}$, and the spectral decomposition is\begin{equation}
\hat{{\cal{H}}}=\int \lambda d\hat{{\cal{H}}}(\lambda)\;.\label{eq:Sch2}
\end{equation}
From the elementary property of the spectral components $\hat{{\cal{H}}}(\lambda)$ in Eq.(\ref{eq:Sch2}),
\begin{equation}
\hat{{\cal{H}}}({\lambda_1})\hat{{\cal{H}}}({\lambda_2})=\delta_{\lambda_1,\lambda_2}\hat{{\cal{H}}}({\lambda_1})\;,
\end{equation}
it follows that,
\begin{equation}
\hat{{\cal{H}}}^2=\int \lambda^2d\hat{{\cal{H}}}(\lambda)\;,
\end{equation}
and the time development in Eq.(\ref{eq:est}) satisfies the properties of a contraction semigroup in the parameter $t$.

The d.o.f. of collapses of the superposition of wave functions \begin{equation}\psi=\sum_\lambda c_\lambda \psi^\lambda\;,\label{eq:super3}\end{equation} is the spectral component $\hat{{\cal{H}}}(\lambda)$. In the superposition of Eq.(\ref{eq:super3}), each component $\psi^\lambda$ is distinguished from the others by the spectral components $\hat{{\cal{H}}}(\lambda)$ such that \begin{equation}{\mbox{if}}\ \ \hat{{\cal{H}}}(\lambda)\psi^{\lambda_1}\neq0\;,\ \ {\mbox{then}}\ \ \hat{{\cal{H}}}(\lambda)\psi^{\lambda_2}=0\;,\end{equation} for $\lambda_1\neq \lambda_2$ and the state space ${V}$ of the system. Concretely, the spectral component $\hat{{\cal{H}}}(\lambda)$ is defined by the restriction of $\hat{{\cal{H}}}$ on the part which lies within the eigenspace $V_\lambda$ for eigenvalue $\lambda$,\begin{equation}\hat{{\cal{H}}}(\lambda)=\hat{{\cal{H}}}|_{V_\lambda}\;,\ \ V=\bigoplus_\lambda V_\lambda\;,\end{equation} which induces a non-unitary action on the wave function within the non-zero quantum variance of the increment of time as an operator of the contraction semigroup in the time evolution.
 We identify the cause of the state reduction with this non-unitary action on the wave function. 
 
 As an easy but a very important remark, due to Eq.(\ref{eq:est}), the decoherence time tends to zero for the macroscopic objects.

 \end{appendix}

\end{document}